\documentclass[%
 aip,
 jmp,%
 amsmath,amssymb,
 reprint,%
]{revtex4-1}

\usepackage{graphicx}
\usepackage{dcolumn}
\usepackage{bm}
\usepackage{float}
\usepackage{hyperref}

\begin{document}


\title[Loss of transmission on directed networks due to link deletion]{Loss of transmission on directed networks due to link deletion}

\author{G. Kashyap}
\affiliation{Indian Institute of Science Education and Research (IISER) Pune, Pune - 411008, India}
\author{G. Ambika}
\email{g.ambika@iisertirupati.ac.in}
\affiliation{Indian Institute of Science Education and Research (IISER) Tirupati, Tirupati - 517507, India}


\date{\today}

\begin{abstract}
We investigate the effects of link deletion on a 1 - to - 1 transmission process occurring on various models of directed networks and characterize them in terms of the fraction of successful transmissions and average transmission time. We make intuitive arguments and numerically show that the probability of successful transmission depends, on an average, on the fractional size of the Strongly Connected Component (SCC) and the average transmission time depends on the Average Path Length (APL) of the SCC. In the context of a specific road-network, we study the roles of various process parameters. Finally, we provide remedial strategies to improve transmission during link deletion in such networks.
\end{abstract}

\pacs{Valid PACS appear here}
\maketitle

\begin{quotation}
Our work explores the relatively less studied problem of robustness in directed networks.
We focus on the structural response of directed networks to random and targeted link deletion and the resulting effects on an abstract 1-to-1 transmission process occurring on the same network. The changes in the network structure are quantified using the relative size of the strongly connected component and its average path-length while the transmission process is measured in terms of successful transmissions and transmission-times. Using a road-network, based on traffic-flow between road-intersections, we explore the role of process parameters like request-rate, probability of deletion and probability of transmission. Unlike many studies that focus on the robustness of transport networks in terms congestion and overloading, the present work implements actual movement between designated origin and destination nodes via pre-assigned routes. We also propose a method of weighted selection of nodes to partially counteract the effects of link deletion and consequent interruption of transmission.

\end{quotation}

\section{\label{sec:s1}Introduction}

Complex networks have become an integral part of any systemic analysis. Their ability to model the essential elements of a system of interacting entities makes them a versatile tool with immense scope for applications \cite{strogatz2001exploring,boccaletti2006complex,newman2003structure,albert2002statistical}. The framework of networks has found relevance in physical, biological \cite{alm2003biological,alon2003biological,pavlopoulos2011using,mason2007graph} and chemical systems \cite{wagner2001small,jeong2000large}, transport systems \cite{guimera2005worldwide,yang1998models,xie2007measuring}, social systems \cite{kumar2010structure,mislove2007measurement,leskovec2008microscopic}, medicine \cite{barabasi2011network,barabasi2007network}, finance and risk \cite{gai2010contagion,acemoglu2015systemic} etc. Depending on the problem at hand, they have been used for visualization, robustness/vulnerability analysis, community-detection, link-prediction, information-spreading etc.

While there is no dearth of problems to be studied on a network, of particular interest to us is the structural response of a network to loss of links and the consequent effects on a transmission process occurring on the same network. We focus on directed networks, mainly because of their relevance and the obvious gaps in existing literature \cite{fang2014modeling,yan2009random}. In an earlier work \cite{kashyap2019link}, we have dealt with only the structural aspects during random and targeted loss of links. In this work, we introduce a dynamical process taking place on the network, as 1-to-1 transmission with rerouting. 

The 1-to-1 transmission process considered is relevant since it acts as a generalized procedure to study particular kinds of flows in a network and is significantly different from most existing methods used to study spreading processes\cite{colizza2007reaction,kitsak2010identification,nowzari2016analysis,moreno2004dynamics,pastor2015epidemic}. While being implemented as a model of ping generation and propagation, it could represent many applications depending on interpretation with/without relevant modifications. For example, the pings could represent data-packets travelling between routers on the internet or they could be pulses of current passing through transmission lines between input/output gates in a digital circuit \cite{oshida2006packet,teuscher2007nature}. In the case of reaction networks \cite{kaiser2004edge,barkai1997robustness}, the pings are compounds, that are initially reactants but follow a specified sequence of steps to transform into products. 

However, the most relevant application in this context, is the study of traffic flow in transportation (road) networks\cite{sullivan2010identifying,knoop2007approach,herty2003modeling,coclite2005traffic}. In this case, the pings represent traffic (people/vehicles), moving from origin to destination locations. While substantial work has been done on transportation networks and their vulnerabilities and robust design, the vast majority of it uses an approach based on congestion-analysis caused by overloading of nodes/edges \cite{zhao2005onset,holme2003congestion,yan2006efficient,danila2006optimal}. To the best of our knowledge, the proposed kind of study has not been undertaken, especially using directed networks. We hope that this study will lead to a better understanding of the underlying relationship between the structure of the network and the dynamical processes occurring on it and their mutual influences.

\begin{figure*}[ht]
\centering
\includegraphics[width=0.75\textwidth]{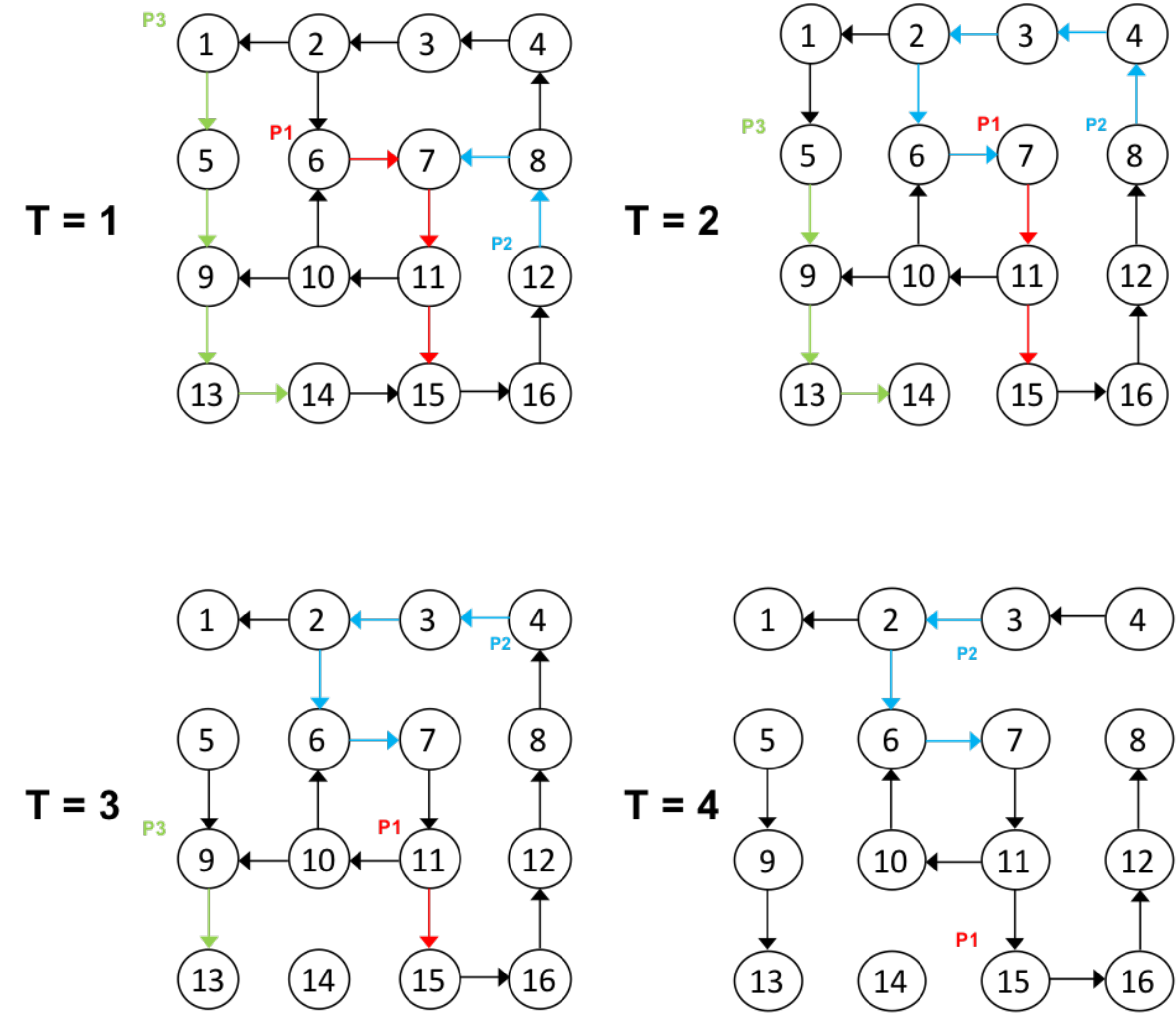}
\caption{Schematic representation of the co-occurring processes of link deletion and transmission. At $T=1$, three pings (P1 - red, P2 - blue and P3 - green) are generated and the pre-assigned paths from start to stop nodes are highlighted in the corresponding colors. At $T=2$, two links are deleted at random and in the process, P2 gets rerouted via a new longer path. At $T=3$, after two more links are deleted, the path of P3 gets disrupted and no alternate path is available and therefore, it is a failed transmission. At $T=4$, P1 reaches its destination and is a successful transmission.}
\label{fig:Schematic}
\end{figure*}

This rest of this paper is organized as follows: we introduce the two processes, define the various parameters and give their relevance in sec - \ref{sec:s2}. In sec - \ref{sec:s3}, we discuss results of co-occurring processes of transmission and link deletion in standard network models and also the role of degree-correlations. In the following section, we bring out the relationship between the processes via qualitative arguments and the correlation between their respective quantifiers. Using data from a specific road-network, in sec - \ref{sec:s5}, we analyze the independent impact of three process parameters. Finally, we discuss the results of a possible strategy, for mitigating the effects of deletion, in sec - \ref{sec:s6}.

\section{\label{sec:s2}Co-occurring processes}

We begin by presenting the finer details of the individual processes, link deletion and 1-to-1 transmission, and provide all the required terminology and the relevant definitions. We start with a discussion of the link deletion process and then move onto the transmission process. This helps us to bring out the unidirectional influence of the former on the latter. A schematic depiction of the processes is shown in fig - \ref{fig:Schematic}.

\subsection{\label{sec:s21}Link Deletion}

The deletion process is controlled by two important parameters: rate of deletion ($v_D$), which is defined as the maximum number of links that can be deleted at any given time-step and the probability of deletion ($p_D$) that is the probability that a selected link is actually deleted.

A predetermined number of links, $v_D$, are selected at every time-step. This selection is based on one of the three strategies of deletion, as described in our earlier work \cite{kashyap2019link}. In the first strategy, labelled S1, we select $v_D$ links uniformly at random from the edge-list and delete each of them with a probability $p_D$. The second and third strategies are methods of targeted deletion. In the second strategy (S2), each link is assigned a weight equal to its edge betweenness centrality(EBC). At each time-step, $v_D$ links with the highest values of EBC are selected and each of them is deleted with a probability $p_D$. The third strategy (S3) is similar to S2 but the links are weighted with values of edge-degree (ED).

Every instance of deletion results in certain structural changes in the networks. To quantify these changes, we calculate the fractional size of the largest strongly connected component (SCC) and the average path-length in the SCC. The extent of connectivity in the network is captured by the size of the SCC and the average path-length of the SCC estimates the efficiency of connectivity. In sec - \ref{sec:s4}, we explain the reasons for evaluating the path-length of the SCC rather than that of the entire network. We also point out that the deletion process depends solely on the structural properties of the network and does not depend on the transmission process in anyway.

\subsection{\label{sec:s22}1-to-1 transmission}

The 1-to-1 transmission process occurs simultaneously with the deletion process. A fixed number of requests/pings, given by the request-rate ($r$), are generated at every time-step $t$. Each ping is associated with randomly chosen start and stop nodes. Based on the network structure at $t$, the shortest (most efficient) path from the start-node to the stop-node is identified and assigned to the respective pings. A ping is characterized as a $failed$ transmission and terminated if no valid path can be found. If multiple shortest paths are available for a ping, then one of the paths is chosen at random. Pre-assigning of specified paths to the pings, sets this study apart from those based on the random-walk models. Once a valid path has been assigned, at every time-step, the pings can travel a maximum distance, specified by the value of \textit{transmission velocity} ($v_T$), with a \textit{transmission probability} $p_T$. Introducing a separate probability for transmission effectively models the traversal of fractional links.

Due to the process of link deletion happening simultaneously in the network, for any ping, the assigned path may be disrupted. In such a case, when the ping arrives at the deleted link, it is rerouted from its current location to the stop-node via a new shortest path. If such a path cannot be found, the ping is deemed a failed transmission. If the ping eventually reaches its destination, either by the initially assigned path or by a rerouted path, it is considered a $successful$ transmission. 

\textit{Ping-time} is defined as the total time taken to travel from the start-node to the stop-node and is only valid for successful pings. It is calculated as the difference between the time a ping is generated and the time it reaches the stop-node. The ratio of the predicted ping-time to the actual ping-time is defined as the \textit{transmission efficiency} and it gives an estimate of the extent of rerouting and thus the effect of deletion on transmission. 

\section{\label{sec:s3}Results on standard network models}

In this section, we present results on networks from standard models such as directed Erd\"os-R\'enyi (ER) networks, directed scale-free (SF) networks and directed small-world (SW) networks.

The process starts with a network size $N$ $(= 1000)$ and average degree equal to 4. At every time-step, links are deleted according to a given strategy. Simultaneously, new pings are generated while existing pings traverse their respective prescribed routes. This sequence of events continues until the number of links remaining is equal to or less than half the number of nodes in the network. The effect of link deletion on the transmission process is characterized in terms of the number of successful pings and the normalized ping times. All results are averaged over 200 realizations. The process parameters $v_D$, $p_D$, $v_T$ and $p_T$ are all set to $1$ and the \textit{request rate} at $r = 5$.

\subsection{\label{sec:s31}Role of topology and strategy}

\begin{figure*}[ht]
\centering
\includegraphics[width=\textwidth]{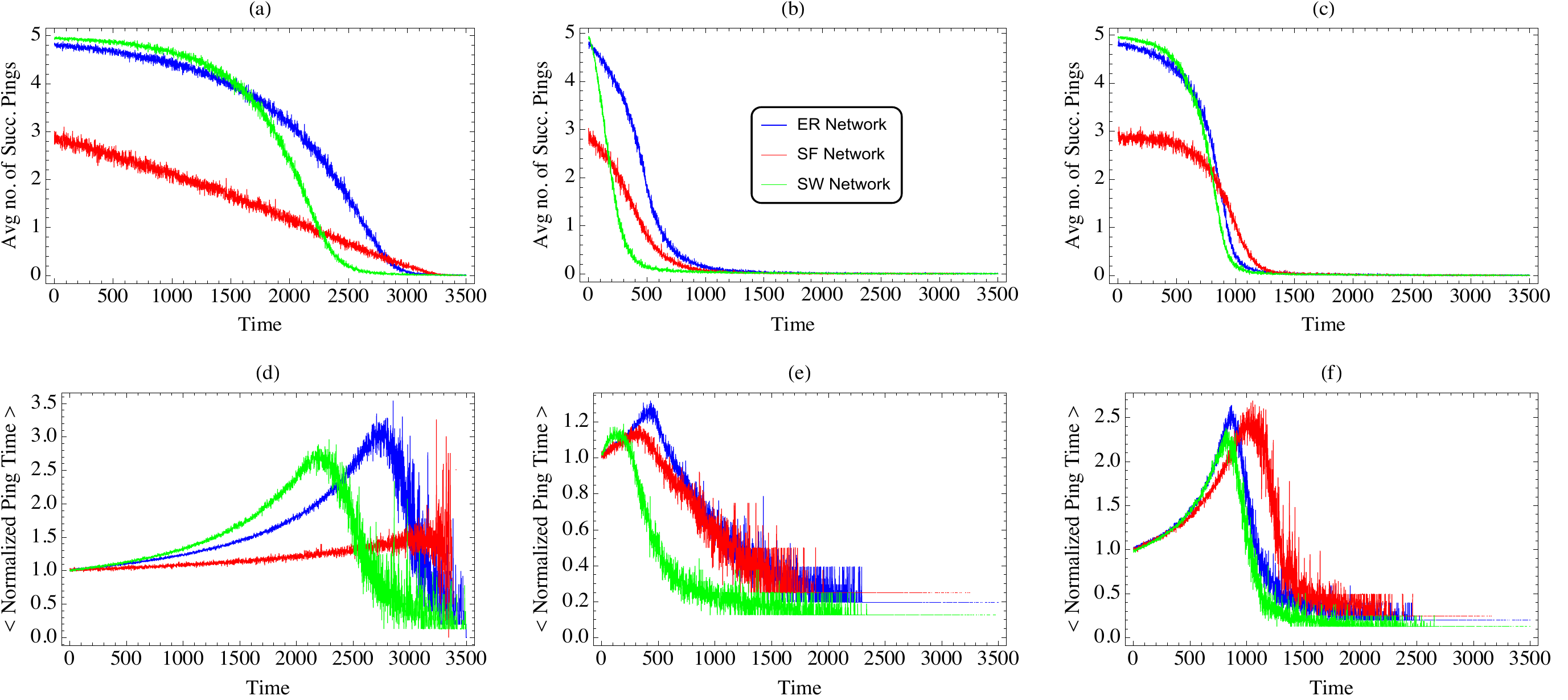}
\caption{(color online) \textit{Top} - Decrease in the average number of successful pings with time (or equivalently loss of links) in ER networks (blue), SF networks (red) and SW networks (green). \textit{Bottom} - The change in behaviour of ping-times during loss of links in the 3 types of directed networks models. The above quantities are studied for S1 - random deletion (a, d), S2 - EBC based deletion (b, e) and S3 - ED based deletion (c, f).}
\label{fig:RoleOfTopology}
\end{figure*}

In the initial stages of random deletion, we observe that the transmission process displays similar behaviour in both ER and SF networks. This is true for both the average number of successful pings and the normalized ping-times. Until about half the initial number of links are deleted, we see a relatively large number of successful transmissions. Beyond this point, in SW networks, the transmission fails very rapidly. Even the normalized ping-times increase exponentially, with a higher rate of change in SW networks as compared to other topologies.

Based on results from earlier study \cite{kashyap2019link}, we know that the process of deletion brings about two types of changes in the network. First, there is a monotonic increase in the path-length within connected sets of nodes and the second change is the appearance of disconnected nodes. This implies that the increase in ping-times must be due to the increase in path-lengths. This also indicates that, while both ER and SW networks are able to sustain a large number of successful pings, the pings are traversing longer distances in SW networks in order to be successful. In the later stages of the deletion process, the behaviour of ping-times also shows a qualitative change. While the initial exponential increase in ping-times is indicative of the dominance of increasing path-lengths, the exponential decrease in ping-times reflects the role of disconnected nodes. Therefore, we deduce that the peak, which marks the change in behaviour, must be the point where the transmission process breaks down. Meanwhile, in SF networks, while there is an overall smaller number of successful transmissions, the behaviour is still linear until breakdown. The role of increased path-lengths and that of rerouting is captured by the behaviour of ping-times, which remain close to 1.

During centrality based deletion, we find SW networks are heavily affected. There is an exponential decrease in the average number of successful transmissions from the initial time-steps, in contrast to ER and SF networks, which have an initial linear decrease that later becomes exponential. Even the normalized ping-times exhibit an exponential drop, indicating a weak influence of longer path-lengths and implying an early abundance of isolated nodes. 

When links are deleted based on ED values (S3), we do not observe any effect of network topology on either the average number of successful pings or the normalized ping-times. In all topologies, we see a sustained number of successful transmissions in the early stages followed by a rapid decrease at almost the same time in all topologies. Such a uniform behaviour is also observed in the case of ping-times, which shows similar levels of increase in all topologies. Overall, it appears that the network topology does not play any role during S3 type of deletion. 

\subsection{\label{sec:s32}Role of degree-correlations}

\begin{figure*}[ht]
\centering
\includegraphics[width=\textwidth]{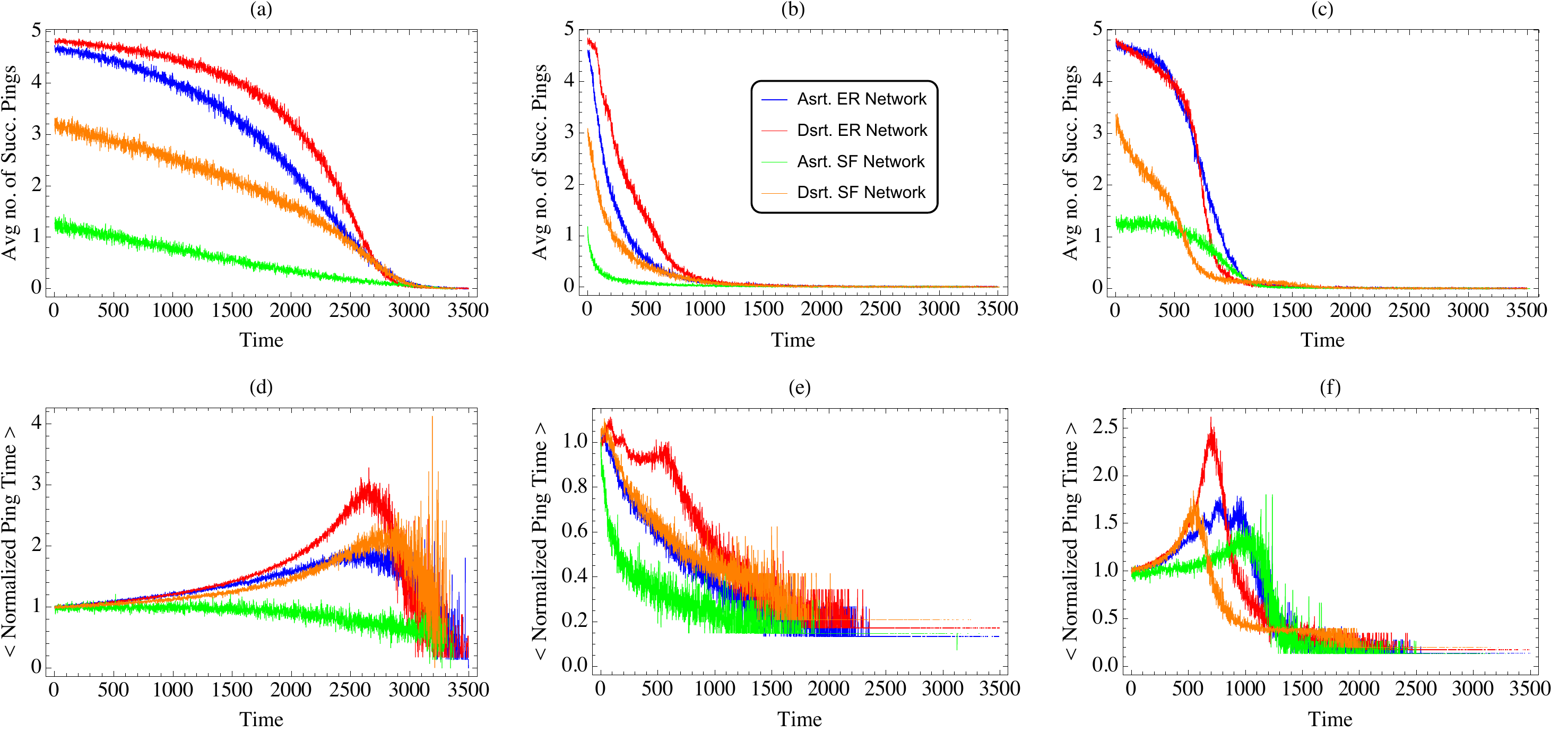}
\caption{(color online) Changes in average number of successful pings/transmissions (top) and average ping-times (bottom) during S1 - (a, d), S2 - (b, e) and S3 - (c, f) types of deletion strategies in assortatively and dissortatively rewired ER and SF networks.}
\label{fig:RoleOfCorr}
\end{figure*}

Next, we study the role of 2-node degree-correlations on the transmission in directed ER and SF networks. We first generate neutral networks with the specified topologies and then tune the extent of correlation using degree preserving rewiring methods, as introduced in \cite{kashyap2017mechanisms}. These networks are then used to study the simultaneous effect of deletion and transmission and the results are presented in fig - \ref{fig:RoleOfCorr}.

In this case, when links are deleted by strategy S1, the results in the case of dissortatively rewired ER networks are comparable to the neutral case. We do not observe any substantial change in the average number of successful pings or the normalized ping-times. In dissortative SF networks, however, we observe a small increase in the number of successful pings but they are less efficient. When ER networks are assortatively rewired, there is considerable decrease in the number of successful pings. These pings are also more efficient as indicated by the smaller rate of increase in ping-times. We observe similar behaviour in SF networks but the increase is not quite significant as in ER networks. 

During S2 type of deletion (EBC based), in both ER and SF networks, there is a decrease in the number of successful pings. This happens during both assortative and dissortative rewiring but the rate of decrease is higher in the case of assortative correlations. The efficiency of transmission is immediately affected in assortative networks by the appearance of isolated nodes, in contrast to the neutral case, where the role of longer, rerouted paths is quite evident from the small initial increase in ping-times. We also observe a small initial resilience to deletion in dissortative ER networks. 

When links are deleted based on ED values (S3), in both assortative and dissortative ER networks, there is no significant effect on the average number of successful pings. However, in the assortative case, the normalized ping-times show a considerable decrease and in the dissortative case, the change is faster as compared to the neutral ER network. In SF networks, there is substantial reduction in the number of successful pings during assortative rewiring but the qualitative behaviour is similar to the neutral SF network. It also results in lower ping-times. When SF networks are rewired for dissortative correlations, there is a linear decrease in successful pings but the behaviour of ping-times is still similar to neutral SF networks.

\section{\label{sec:s4}Relationship between link deletion and transmission}

\begin{figure*}[ht]
\centering
\includegraphics[width=\textwidth]{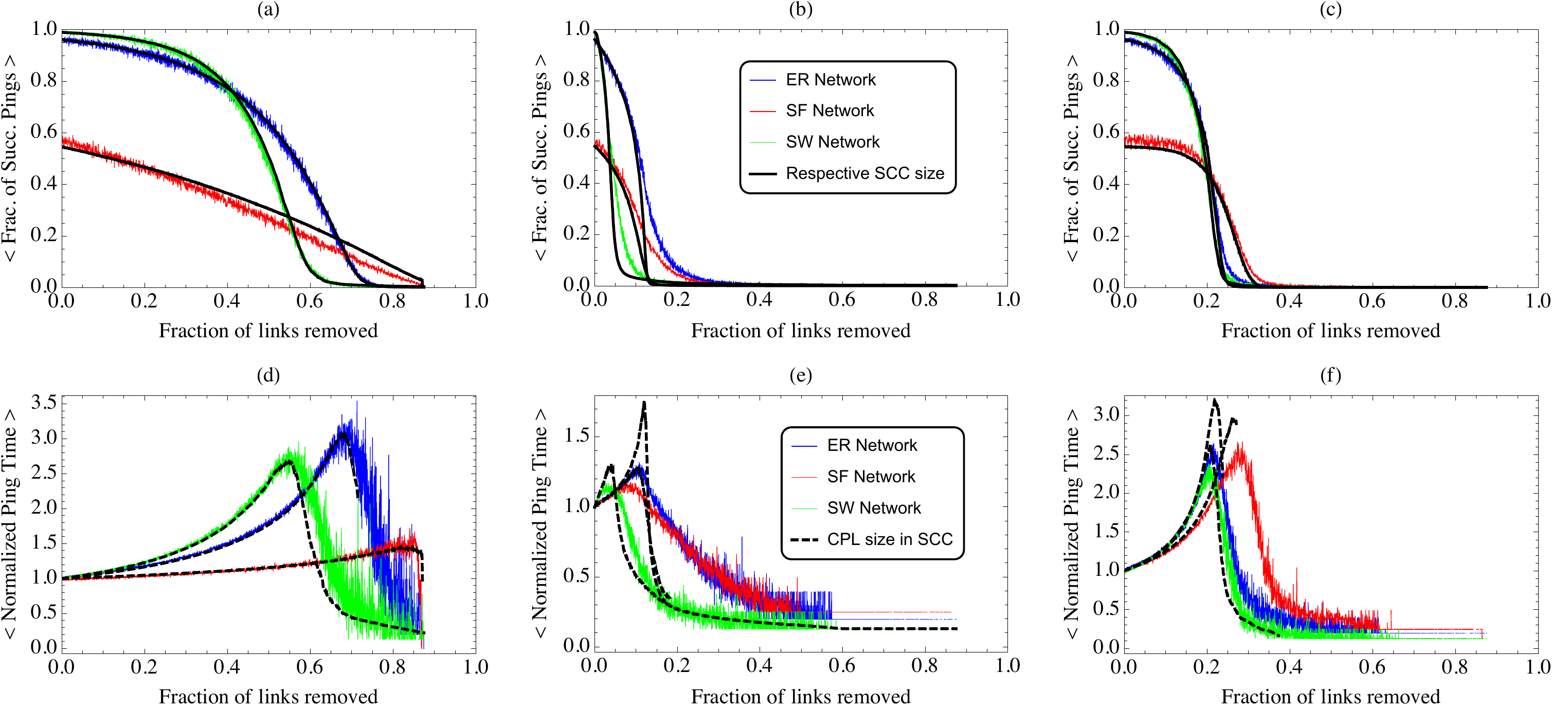}
\caption{Relationship between the link deletion process and the transmission process is explored as the correlation between the average fraction of successful pings and fractional SCC size of the network when plotted as a function of fraction of deleted links (top). We also see that the average path-length of the SCC captures the behaviour of average ping-times as the process of deletion progresses (bottom). This observation remains consistent for all network topology and strategy of deletion.}
\label{fig:NormalizedPlots}
\end{figure*}

In our study, we have the process of link deletion, which directly affects the network structure and the transmission process, which is directly affected by the network structure. Since both processes occur simultaneously and can be parameterized by the same time variable, we can study the transmission process as a function of the deletion process. This helps us to better understand the relationship between the two processes. To this end, in fig - \ref{fig:NormalizedPlots}, we plot the fraction of successful pings and the normalized ping-times against the fraction of link deleted, which is an indicator of the connectivity in the network.

First, we look at how the deletion process affects the transmission process via the size of the Strongly Connected Component (SCC). For any given ping, successful transmission is possible only if there exists a path from the start-node to the stop-node. Such a path need not be unique but must be available, subject to rerouting, until the stop-node is reached. However, a path in the opposite direction is not necessary. When interpreted in the context of a network, this condition requires the start and stop nodes to be part of the Weakly Connected Component (WCC). Therefore, the probability that a ping is successful appears to depend only on the fractional size of the WCC and not on the SCC. Now, if we consider a large number of such pings generated at a given time-step, where the start and stop nodes are chosen uniformly randomly, the probability of successful transmission depends on the existence of unique paths in both to and fro directions. This is because it is possible to have two unique pings for which the start and stop nodes are simply interchanged. This requires us to account for unique paths in both directions. In other words, it is necessary to account for paths between all unique pairs of nodes in the network. This implies that the probability of success depends on the probability that the start and stop nodes belong to the SCC and consequently on the fractional size of the SCC itself. Therefore, the fractional size of the SCC is an indicator of the average behaviour of the probability of successful transmission. This holds true independent of the network topology and strategy of deletion. 

At time $t$, let $R$ be the number of pings generated and $S$ be the number of successful pings, then the probability of successful transmission at $t$ is given by

\begin{equation}
    \label{eqn:e1}
    \frac{S}{R} = f(t)
\end{equation}
where $f$ is the time-dependent probability and is equal to the fractional size of the SCC. If we consider the pings generated in an interval of time $\Delta t$, then we have $S/R=f(\Delta t)$ and $f$ is the fractional size of the SCC that remains constant in $\Delta t$. In terms of the $request$ $rate$ $r$, eqn - \ref{eqn:e1} can then be written as 

\begin{equation}
    \label{eqn:e2}
    S = f r \Delta t
\end{equation}

By extending this line of argument, we can also deduce the relationship between the average path-length of the SCC and the average ping-times. Since, on average, successful pings are confined to the SCC, it is only logical that the characteristic length-scale of the SCC must affect the ping-times. Indeed, we find that the behaviour of the normalized ping-times is captured by the average path of the SCC. This is only true on average because it possible for some successful pings leading into and out of the SCC. There is also an advantage because the path-length of the SCC remains well-defined during the deletion process in contrast to the path-length of the whole network. This result also does not depend on the topology or the strategy of deletion.

For a ping generated at time $t$ and travelling a distance $d_p$, let the time taken be $t_p$. Then

\begin{equation}
    \label{eqn:e3}
    t_p = \frac{d_p}{p_T * v_T}
\end{equation}
If we consider many such pings at $t$, and given that $p_T$ and $v_T$ are fixed, then the average ping-time is given by

\begin{equation}
    \label{eqn:e4}
    <t_p> = \frac{<d_p>}{p_T . v_T} = \frac{<APL_{scc}>}{p_T . v_T}
\end{equation}

\section{\label{sec:s5}Role of process parameters - Barcelona road-network}

In this section, we present the analysis on the independent roles of the different process parameters. Although we have $v_D$ and $p_D$, associated with the deletion process and $v_T$, $p_T$ and $r$, associated with the transmission process, we limit our study to the roles of $r$, $p_T$ and $p_D$. This is because these three parameters can capture the essential dependencies while the remaining parameters can provide for greater control during implementation. We can also avoid specifically dealing with network related properties like size, density etc. because the measures are normalized accordingly.

In this section, instead of considering standard network models, we focus on a road-network based on the city of Barcelona, where the nodes represent road intersections and the flow of traffic between the intersections constitute the links. The network consists of 1020 nodes and 2522 links and the largest connected component spans 992 nodes. It has a lattice-like structure with small positive degree correlations and negligible clustering. 

\begin{figure*}[ht]
\centering
\includegraphics[width=\textwidth]{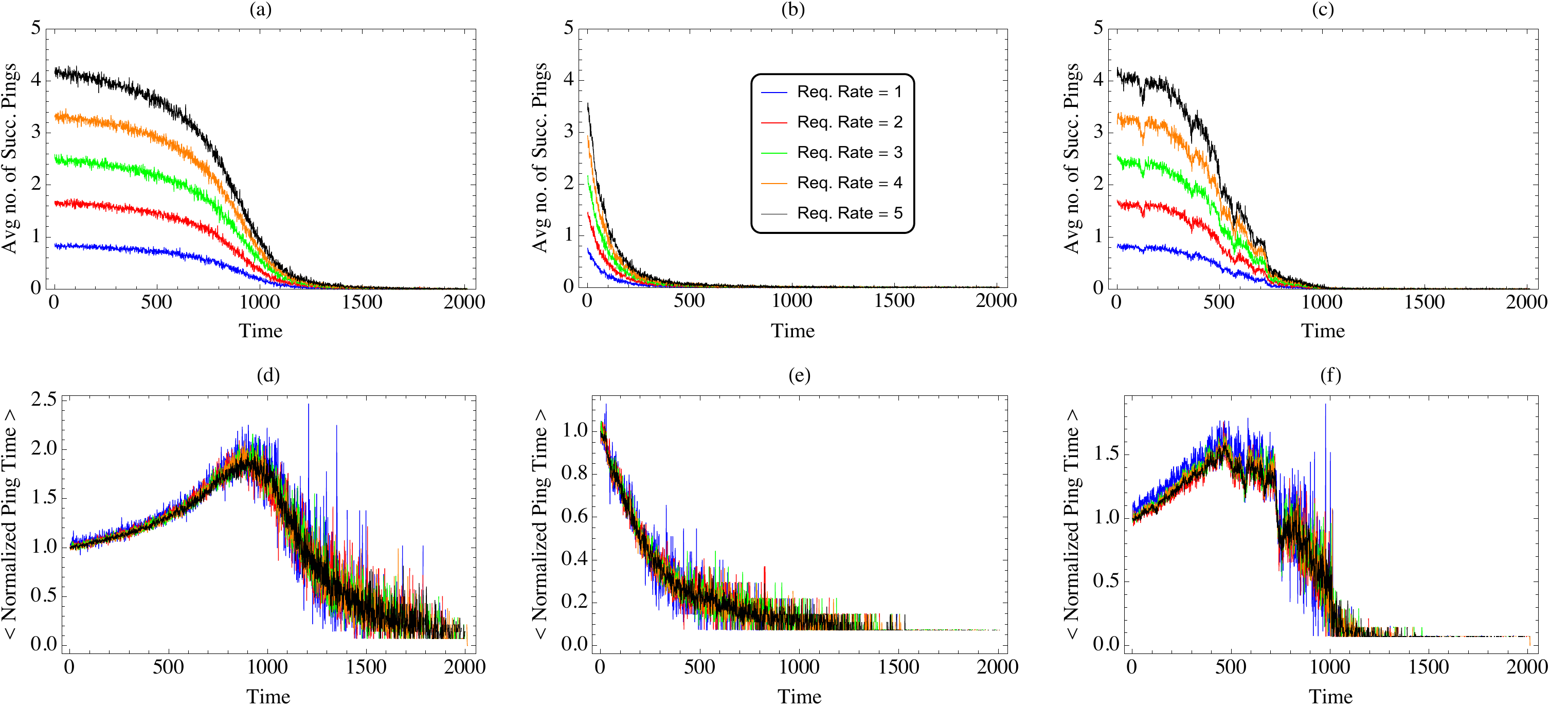}
\caption{(color online) Effect of changing request rate on the average number of successful pings (top) and the average normalized ping-times (bottom) during link deletion by various strategies (S1 - a, d; S2- b, e; S3 - c, f) in the Barcelona road network. Plot as time-series}
\label{fig:RoleOfReqRate}
\end{figure*}

We recast the problem of 1-to-1 transmission to model the flow of road-traffic in an arbitrary geographic setting. In this particular problem, each trip served by a vehicle has unique origin and destination locations. As the nodes in the network are distinctly defined, these locations can be approximated to the nearest intersections/nodes. The total distance travelled in a trip, including fractional links, is calculated as the product of $v_T$ and $p_T$. A deleted link is equivalent to an inaccessible road. We do not specify how or why a road becomes inaccessible. This could happen due to many reasons such as accidents, traffic-jams, repair-works etc., which could broadly be classified as either systematic or random. When faced with an inaccessible road, the vehicle has a choice to either cancel the trip, which is equivalent to a failed transmission, or take a less efficient alternate route (rerouting). The cancelling of trips, although not a very realistic scenario, is implemented to stay consistent with the fact that deletion is irreversible.

\subsection{\label{sec:s51}Role of request-rate}

The average number of pings/trips generated in unit time is defined as the \textit{request rate} $r$ and the trips starting at time $t$ are represented by $r(t)$. We simplify the model so that there is no constraint on the number of vehicles at any node or link. As a result, if there is an increase in the number of trips starting at $t$, we should observe a proportional increase in the number of successful trips and the probability of a successful trip should remain fixed for constant $r$.

For unit time interval, we see from eqn - \ref{eqn:e2} that $S(t)$ is proportional to $r$. We also know that $S(t)$ is related only to the transmission process and $f(t)$ is related only to the deletion process implicitly through the surviving fraction of links $m(t)$. For a single instance of deletion, we let $r$ take a fixed integer value or draw a value from a Poisson distribution with fixed mean. While the second option is more realistic, it does not provide any additional understanding of the process. We see that the dynamical behaviour of $S(t)$ is dependent only on $f(t)$ since any change in $r$ does not affect the dynamics of the transmission process. However, at a specified value of $t$, $S(t)$ is multiplied by a factor $r$. These results are shown in fig - \ref{fig:RoleOfReqRate} (a - c).

As the process of deletion progresses, the behaviour of $f$, as a function of fraction of remaining links ($m$) is specified by the type of deletion strategy. The role of $r$ is itself independent of the deletion strategy. For $p_D = v_D = 1$, the behaviour of $f(m)$ controls the qualitative behaviour of the slope of $S$ while the exact value at any $t$ is scaled by the value of $r$. Therefore, from fig - \ref{fig:RoleOfReqRate}, we see that the rate of decrease of $S$ is proportionally higher for higher values of $r$. When it comes to the normalized ping-times, eqn - \ref{eqn:e4} tells us that the ping-times are independent of $r$. We find that this is indeed the case, as shown in fig - \ref{fig:RoleOfReqRate} (d - f).

\subsection{\label{sec:s52}Role of transmission probability}

\begin{figure*}[ht]
\centering
\includegraphics[width=\textwidth]{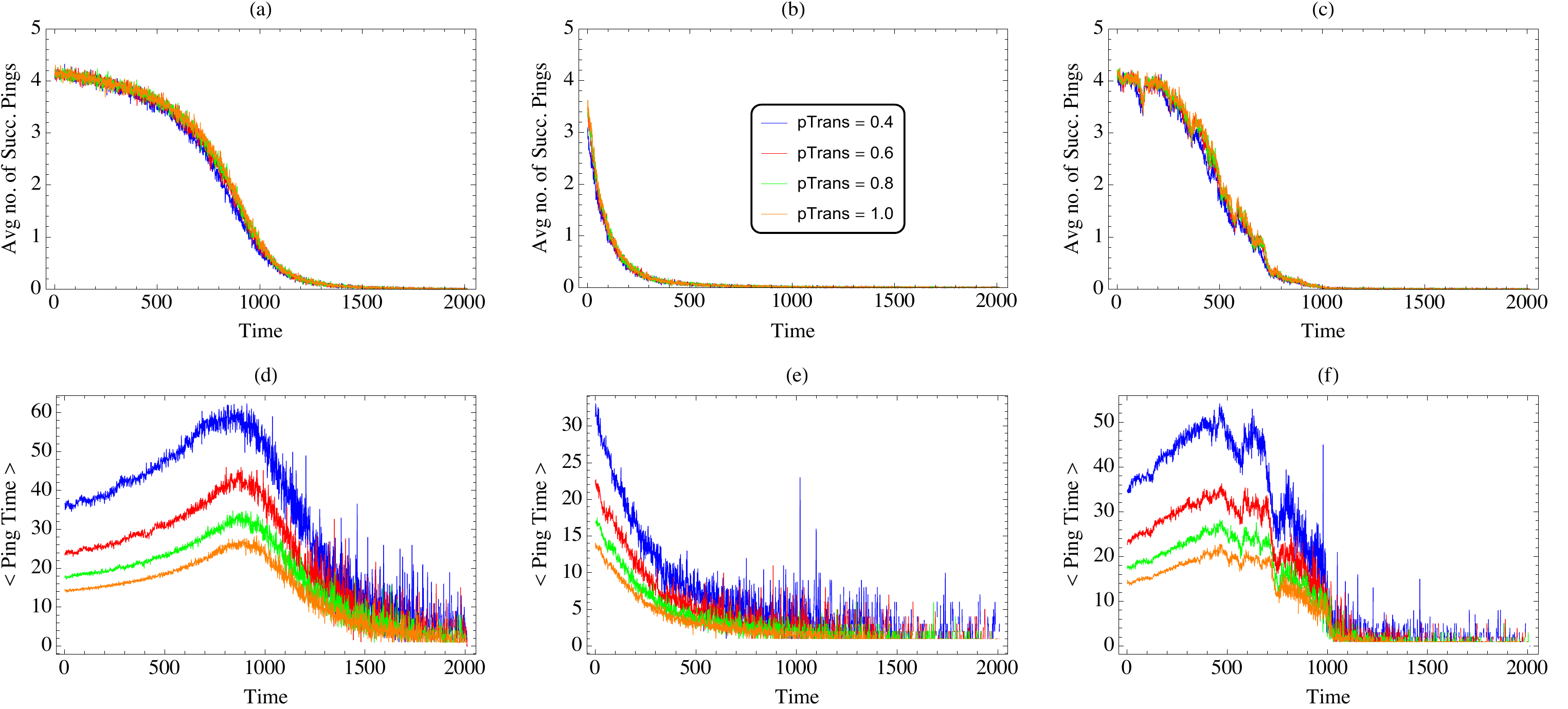}
\caption{(color online) Effect of decreased probability of transmission on the average number of successful pings (top) and the average normalized ping-times (bottom) during link deletion by various strategies (S1 - a, d; S2- b, e; S3 - c, f) in the Barcelona road network.}
\label{fig:RoleOfpTrans}
\end{figure*}

During the process of transmission, each ping traverses at most $v_T$ links in a given time-step, each with a probability $p_T$. Therefore, $p_T$ effectively sets a time-scale for the transmission process w.r.t the deletion process. A value of $p_T$ less than 1 essentially models the traversal of fractional links. Having $p_T$ less than $p_D$ is not a very realistic scenario, especially in the context of traffic-flow on road networks. It points to a possible case of rapidly cascading traffic-jams. It can also be interpreted as a shrinking of the time-interval in which $f$ remains constant. The interval decreases by a factor of $p_T$ and eqn - \ref{eqn:e2} becomes 

\begin{equation}
    \label{eqn:e10}
    S = fr.(p_T . \Delta t)
\end{equation}

From the above equation, we see that the network loses links faster than the pings can traverse them and this results in a decrease in the probability of successful trips. Due to the effect of rerouting, when the results are plotted as shown in fig - \ref{fig:RoleOfpTrans} (a - c), the role of transmission probability gets masked and is not immediately obvious.
Given $p_T < 1$, for any trip starting at $t$, the assigned path has a greater chance of disruption due to link deletion. This does not immediately result in loss of transmission but gets compensated by an alternate path. The process of rerouting taps into the large number of redundant paths present in the network. As a result, the average number of successful trips is sustained but the efficiency of travel is negatively affected. 

The probability of transmission plays an important role in the response of ping-times to link deletion. Values of $<t_p>$ get scaled by a factor of $1/p_T$, as derived from eqn - \ref{eqn:e4}, and shown in fig - \ref{fig:RoleOfpTrans} (d - f). Our explanation, that trips get rerouted earlier and take longer paths, gets additional support from this observation. We also find that there is no effect of $p_T$ on the rate of change of $<t_p>$ and this is because the time-derivative of eqn - \ref{eqn:e4} is inversely related to $p_T$. Pings generated at $t$ travel a distance, that increases proportional to $p_T$, and therefore the rate of change of $<t_p>$ appears to be independent of $p_T$. As a result, the ping-times would overlap and their dependence would not be obvious. Therefore, we discard the normalized values and indicate the actual ping-times.

\subsection{\label{sec:s53}Role of deletion probability}

\begin{figure*}[ht]
\centering
\includegraphics[width=\textwidth]{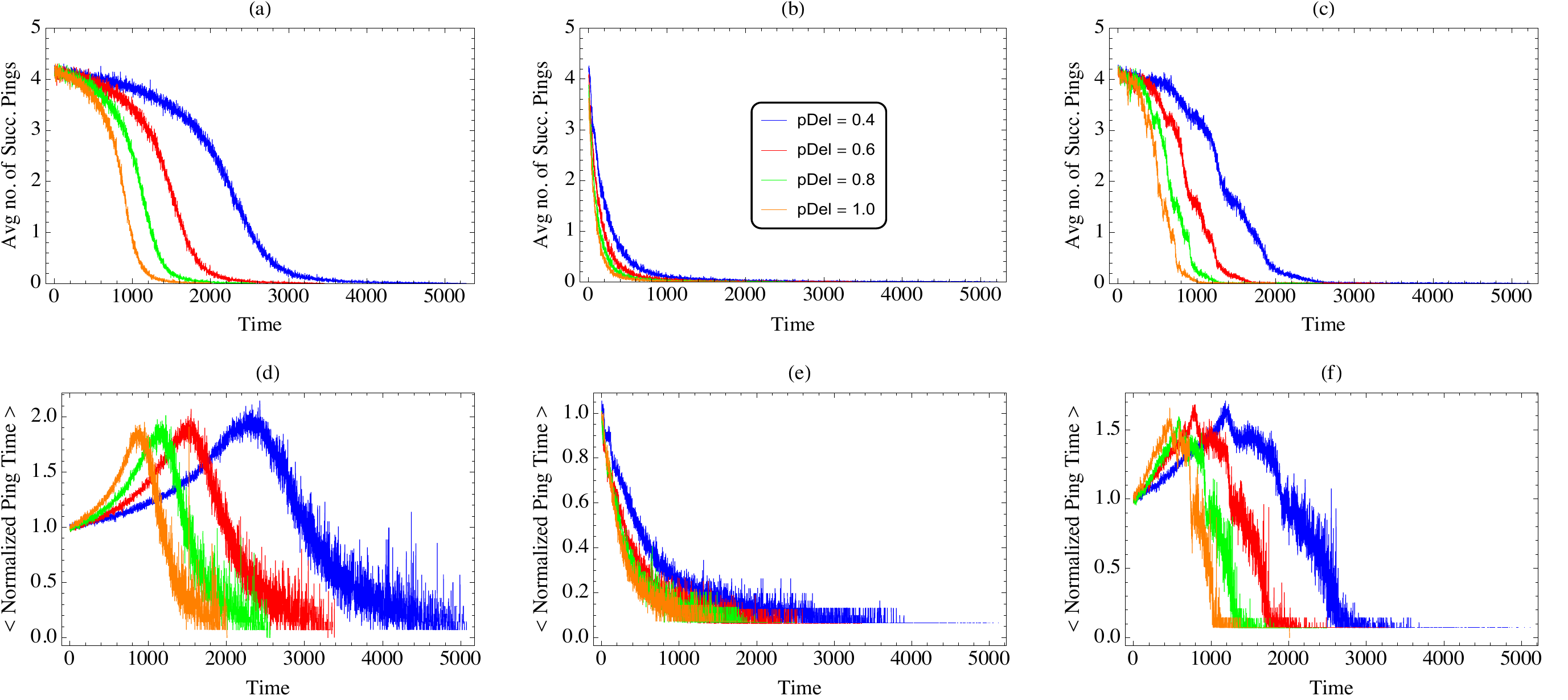}
\caption{(color online) Effect of decreased probability of deletion on the average number of successful pings (top) and the average normalized ping-times (bottom) during link deletion by various strategies (S1 - a, d; S2 - b, e; S3 - c, f) in the Barcelona road network.}
\label{fig:RoleOfpDel}
\end{figure*}

At any given time-step, only a fraction of the total number of selected links ($v_D$) are deleted. Each selected link is deleted with a probability of deletion $p_D$. This probability $p_D$ also sets the time-scale for the deletion process relative to the transmission process. Since a value of $p_D < 1$ implies that only a fraction $p_D$ of $v_D$ selected links are deleted, in terms of the time-scales, it implies that the time-interval required to delete $v_D$ links is increased by a factor of $1/p_D$. Eqn - \ref{eqn:e2} then becomes 

\begin{align}
    \begin{split}
        S = \frac{f.r.\Delta t}{p_D}
    \end{split}
    \label{eqn:e6}
\end{align}

The average number of successful trips, at any time-step, increases by a factor of $1/p_D$ (fig - \ref{fig:RoleOfpDel} (a - c)) and this is consistent with eqn - \ref{eqn:e6}. This is true for all strategies of deletion and occurs over the behaviour imposed by the rate of change of $f$. The rate of decrease of $S(t)$ also comes down by a factor of $p_D$ for all deletion strategies. This is shown in fig - \ref{fig:RoleOfpDel} (a - c), for $r=5$ and $v_D=1$.

For values of $p_D < 1$, the interval between deletions become longer and therefore the trips have more time to reach their destinations. However, the distance traversed in each trip is in no way affected. This results in an increase in $S(t)$ at all $t$. Even for trips whose initial paths have been disrupted, the expanded intervals between deletions provide more time to reach their respective destinations via longer rerouted paths. This leads to an overall increase in the average distance travelled at any given $t$ and is reflected in the net increase in normalized ping-times (Fig - \ref{fig:RoleOfpDel} (d - f)). Due to the slow rate of deletion, rerouting of trips, if necessary, is delayed. Therefore, for small values of $p_D$, we see a slow rate of increase in average ping-times. This can also be interpreted from eqn - \ref{eqn:e4}. 

The $APL$ of the SCC depends on the density of links in the SCC ($\overline{k}_{scc}$) and also on the size of the SCC itself. But, numerically, we observe that, before the appearance of isolated nodes, the average degree of the $SCC$ remains quite constant and therefore the rate of change of $\overline{k}_{scc} = 0$.
Since the rate of change of $f$ is a monotonically decreasing function, the overall slope becomes positive and is scaled by a factor $p_D$. This explains the behaviour of normalized ping-times in fig - \ref{fig:RoleOfpDel} (d - f), in the region dominated by increasing path-length. When isolated nodes begin to appear, this argument becomes invalid because the average degree of the $SCC$ does not remain constant anymore.

\section{\label{sec:s6}Strategy for mitigation - Weighted selection}

\begin{figure*}[ht]
\centering
\includegraphics[width=\textwidth]{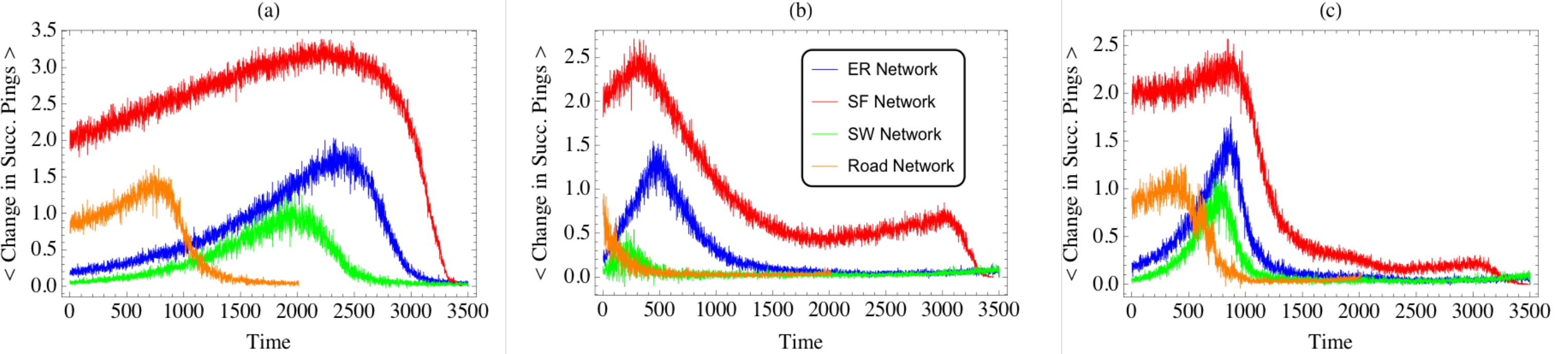}
\caption{(color online) \textit{Top} - Average change in the number of successful pings during link deletion in ER networks (blue), SF networks (red), SW networks (green) and Road network (orange) due to weighted selection of endpoints for individual pings. \textit{Bottom} - The effect of weighted selection of endpoints of pings on the average normalized ping-times in the various networks during various types of link deletion (S1 - a, d; S2 - b, e; S3 - c, f).}
\label{fig:RoleOfWSel}
\end{figure*}

In the final part of the paper, we present a strategy that allows us to at least partially improve the performance of the transmission process. By first analysing the mechanism(s) by which the process of link deletion affects transmission in the network, we build an appropriate procedure to counteract such effects. Using such an approach, we find that the transmission is affected primarily due to loss of shortest paths.

Since shortest-paths play such an important role in sustaining transmission, we can put together a sequence of steps to either increase the number of shortest paths or decrease their average length. There are two ways in which this can be achieved: (a) We select a suitable topology (like SF or SW) in which the average path-length is small, so that there is higher probability of successful transmission and (b) irrespective of the topology, we select a pair of nodes for which the average number of connecting paths is significantly high. In this work, we proceed with the second option and discuss the method of selection, the reason for increased transmission and the limitations of such a method.

In our strategy, we propose to make a weighted selection of start and stop nodes. When a new trip is generated, the start-node is selected with a probability proportional to the out-degree of the nodes and the stop-node is selected with a probability proportional to the in-degree of the nodes. This method of selection becomes very relevant, especially when interpreted in the context of a traffic-flow problem. The probability of a trip between any two given stops is not uniform because of the heterogeneity in importance of the locations. Locations with greater importance become more likely to be associated with a greater fraction of trips. It is also well known that such important locations tend to be better connected. Therefore, if we consider high-degree nodes in the network to be the more important stops, then the model has essentially assigned a larger number of trips within the more important locations. This method of weighted selection works mainly because nodes with high-degrees tend to have better connectivity just by virtue of their high-degree.

Using weighted selection of origin and destination locations, the entire procedure is repeated in all network topologies for all strategies of deletion. The net increase in the average number of successful transmissions, plotted as a function of start-times, is shown in fig - \ref{fig:RoleOfWSel}. During random deletion, this strategy is found to be highly effective and we observe a maximum increase of more than three successful transmissions per time-step. Between strategies S2 and S3, we find that while the maximum increase is more than two successful transmissions per time-step, the increase is better sustained during S3 as compared to S2.

This weighted selection of stops brings about a maximum increase in transmissions in SF, ER, SW and road networks, in that order, and it is independent of the strategy of deletion. Since this method depends on the number of nodes with relatively high degrees, the heterogeneity of the degree distributions plays an important role. As a result, SF networks exhibit the maximum increase in transmissions due to the heavy-tail degree distribution and the road network has the least change because of its lattice-like structure. When the connectivity is mostly homogeneous, there is little distinction between high-degree and low-degree nodes and therefore weighted selection does not differ very much from random selection.

While the proposed method certainly improves the probability of transmission, it is by no means a unique choice. In the current context, an alternative measure would be to assign less efficient paths to the trips when they are generated, so that the trips are not immediately affected by the deletion process. Allowing for early rerouting of trips, rather than holding out until the deleted link is reached, is another viable alternative. Similar procedures can be designed depending on the specific problem at hand. 

\section{\label{sec:s7}Conclusion}

In summary, we study the process of random and targeted link deletion in directed networks and their effects on a 1-to-1 transmission process co-occurring on the same network. The study is broadly divided into two parts. In the first part, we focus on the dynamics of the transmission process in network models. We study the roles of directed network topologies and various strategies of deletion and also the effects of 2-node degree-correlations. In any topology, we observe that maximum disruption is caused by centrality based deletion, followed by degree-based deletion and random deletion respectively. During random deletion, ER networks exhibit very robust transmission. During ED based deletion, however, transmission in all topologies is equally affected. Depending on the topology, positive correlations have a differential effect on the resilience while negative correlations do not appear to play any significant role. By studying the dynamics of the network metrics and the quantifiers of the transmission process, we develop qualitative arguments to build the relationship between the two processes. 

In the second part of our study, the process of 1-to-1 transmission is recast as a problem of traffic-flow on a road network. We use a traffic network, constructed from road-intersection data in the city of Barcelona, to study the role of parameters associated with the transmission process. The results are then interpreted in the context of a traffic-flow problem. The request-rate is found to affect only the average fraction of successful pings and has no effect on the normalized ping-times. On the other hand, the transmission probability has an effect only on the ping-times. We also observe an increase in the probability of success for small values of deletion probability. Finally, we propose a mechanism, based on weighted selection of start and stop nodes, and using qualitative arguments and numerical evidence, we show that it can mitigate the loss of transmission.


%

\end{document}